\documentclass[aps,prb,twocolumn,showpacs,groupedaddress]{revtex4}
\usepackage{graphics,epsfig,comment}
\begin{document}
\title{Accidental Degeneracy in k-space, Geometrical Phase, and the Perturbation
of $\pi$ by Spin-orbit Interactions}
\author{Philip B. Allen}
\affiliation{Department of Physics and Astronomy,
             Stony Brook University, Stony Brook, NY
             11794-3800
	    }
\author{Warren E. Pickett}
\affiliation{Department of Physics, University of California Davis,
   Davis CA 95616}

\date{\today}
\begin{abstract}
Since closed lines of {\it accidental} electronic degeneracies were demonstrated to be
possible, even frequent, by Herring in 1937, no further developments
arose for eight decades. The earliest report of such a nodal loop in
a real material -- aluminum -- is recounted and elaborated on. Nodal
loop semimetals have become a focus of recent activity, with emphasis on
other issues. Band degeneracies are, after all, the origin of topological
phases in crystalline materials.
Spin-orbit interaction lifts accidental band degeneracies, with the resulting 
spectrum being provided here.  The geometric
phase $\gamma(C)=\pm\pi$ for circuits $C$ surrounding a line of such degeneracy
cannot survive completely unchanged.  The change depends on how the spin
is fixed during adiabatic evolution.  For spin fixed along the internal
spin-orbit field, $\gamma(C)$ decreases to zero as the circuit collapses
around the line of lifted degeneracy.  For spin fixed along a perpendicular
axis, the conical intersection persists and $\gamma(C)=\pm\pi$ is unchanged. 
\end{abstract}
\maketitle

\section{Introduction}
A guiding principle of quantum mechanics is that, in the absence of symmetries
that allow crossing of eigenvalues as some parameter in the Hamiltonian is
varied, the eigenstates will repel due to a non-zero matrix elements of the Hamiltonian
and will anticross instead of encountering a degeneracy (eigenvalue crossing).
von Neumann and Wigner explored this question,\cite{Neumann1929} finding that only three
parameters in a Hamiltonian are necessary for an {\it accidental} degeneracy
(one not enabled by symmetry) to occur; only two are required if the
Hamiltonian is real (viz. contains a center of inversion). Bouchaert,
Smolokowski, and Wigner\cite{BSW} 
laid the groundwork for categorizing the symmetry-determined
degeneracies in crystalline solids, which is built on crystal symmetry groups and
subgroups. Symmetry has persisted in being a fundamental organizing principle
in solid state physics. 

Following this development, Wigner gave Conyers Herring the task of investigating
possible accidental degeneracies in crystalline materials. Among Herring's various
findings\cite{Herring1937,HerringThesis} was that 
not only accidental degeneracies will occur in solids, but that closed lines
(loops) of degeneracies are allowed and should not be uncommon. This finding
assumed lack of spin-orbit coupling, which is now well known to (primarily) split
electronic degeneracies, and occasionally to invert the energies of states.
Interest in nodal loops in crystalline spectra has become very active in the
last five years, after 75 years of relative neglect (exceptions include 
Blount\cite{Blount1962}, Zak\cite{Zak}, Mikitik and 
Sharla\v{i}\cite{Mikitik1998,Mikitik1999,Mikitik2007}, and
Allen\cite{Allen2007,PBA-unpubl}).

Separately, a geometric phase in systems undergoing an adiabatic evolution
was identified by Berry,\cite{Berry,Berry1985} 
with characteristics tied to degeneracies. Berry
introduced the term {\it diabolical points}, seemingly because degeneracies
correspond to points in configuration space where the eigensystem suffers a
non-analyticity -- diabolical mathematical behavior. His motivation for this
term was however ascribed to the conical shape of energy spectrum around the
degeneracy point, the shape being that of a diabola involved in the object
manipulated by the toy comprised of sticks and strings. Regardless of the
etymology, diabolic points and geometric phases have subsequently been identified in
numerous systems and occupy a fundamental place in the quantum mechanics of
quasiclassical systems.    

In this paper we begin by providing the prescription for following, once a
degeneracy is located, the degenerate pair around the loop, using modern
notation and presenting algorithms explicitly. Such a loop is shown to carry
a topological index of $\pm \pi$. Then when SOC is included,
as noted in the modern era by Burkov, Hook, and Balents,\cite{Burkov2011} 
the degeneracy is lifted everywhere except
possibly at points where symmetry dictates that matrix elements of the SOC
operator vanish. With the non-analyticity of the eigensystem removed, the
topological index vanishes, and since all materials possess some SOC, the
implication seems to be that the nodal loop is not physical, that it only
existed in a SOC-less universe. Straightforward extension of the same
thinking to an applied magnetic field leads to the result 
that the nodal loop can however still
be located, though the degeneracy never existed. This paper is concluded by
interpreting characteristics of CaAs$_3$, the so-far unique nodal loop semimetal
whose only symmetry is inversion, in terms of this formalism. 

\section{Berry's Formalism}
More specifically for current purposes, the geometric (or Berry) 
phase\cite{Berry} has become a powerful tool for 
analysis of waves in periodic systems, especially
electrons in crystals \cite{Zak,Vanderbilt,Resta,Sundaram,Haldane,Mikitik2007}. 
The wavevector $\vec{k}$ provides a space in which adiabatic evolution of
wavefunctions $\psi_n(\vec{k},\vec{r})$ can be studied, such as by nuclear motion
or by applied fields.  Singular behavior
occurs at band degeneracies where energies $\epsilon_1(\vec{k})=
\epsilon_2(\vec{k})$ are equal and the eigensystem becomes non-analytic.  
In crystals with inversion symmetry,
ignoring spin-orbit interactions, degeneracies occur along closed
lines in $\vec{k}$-space \cite{Blount1962}.  
The periodic part $u_n(\vec{k},\vec{r})=\exp(-i\vec{k}\cdot\vec{r})
\psi_n(\vec{k},\vec{r})$ of $\psi$ is an eigenstate
of 
\begin{eqnarray}
{\cal H}(\vec{k})=(\vec{p}+\hbar\vec{k})^2/2m +V(\vec{r}).
\end{eqnarray}
Let the wavevector $\vec{k}(t)$ be given a time evolution which takes
it on the closed circuit {\it C}, with $\vec{k}(T)=\vec{k}(0)$.

Now suppose that wavefunction evolution
is determined by the time-dependent Schr\"odinger equation with the
time-dependent Hamiltonian ${\cal H}(\vec{k}(t))$.
The time-evolution is assumed adiabatic, namely
$u_n(\vec{k},\vec{r},t) \propto u_n(\vec{k}(t),\vec{r})$.
Berry's argument shows that $u_n(\vec{k},\vec{r},T)$ differs from
$u_n(\vec{k},\vec{r},0)$ by the factor $\exp[i\gamma(C,T)]$, where
the phase $\gamma(C,T)$ has two parts, 
$\gamma(C)+\gamma(T)$.  The dynamical part
$\gamma(T) = -\int_0^T dt \epsilon_n(\vec{k}(t))/\hbar$ depends on the
time elapsed. The geometric part
\begin{equation}
\gamma(C)=i\oint_C d\vec{k}\cdot\int d\vec{r} u_n^{\ast}\vec{\nabla}_k u_n
\label{eq:geomphase}
\end{equation}
is invariant and intrinsic to the circuit and the band properties.
In particular, $\gamma(C)=\pm\pi$ if {\it C} encloses one (or an odd
number) of degeneracy lines.  This change of wavefunction sign is
familiar from other problems where a circuit of adiabatic evolution
surrounds a conical intersection.  However, direct evaluation of 
Eq.(\ref{eq:geomphase}) is problematic, since wavefunctions must be
defined and evaluated in a continuous and single-valued manner.
But when the circuit {\cal C} is discretized in $k$-space for numerical
integration, the code used for $\psi_n(\vec k,\vec r)$ is likely to
produce a random phase $\phi_n(\vec k)_i$ discontinuously jumps 
to a neighboring $\vec k_{i+1}$.

Although gauge invariance is not easy to show in Eq.(\ref{eq:geomphase}),
Berry gave also an alternate form, for a 3-dimensional parameter
space $\vec{k}$, as the flux through a surface {\it S} (bounded by {\it C})
of a vector $\vec{V}_n$.  
\begin{equation}
\gamma(C)=-\int_S d\vec{S}_{\vec{k}}\cdot\vec{V}_n
\label{eq:Sint}
\end{equation}
\begin{equation}
\vec{V}_n = {\rm Im}\sum_m \frac{\langle n|\vec{\nabla}_{\vec{k}}{\cal H}
|m\rangle \times \langle m|\vec{\nabla}_{\vec{k}}{\cal H} |n\rangle}
{[\epsilon_m(\vec{k}) -\epsilon_n(\vec{k})]^2}
\label{eq:V}
\end{equation}
The gauge invariance of this vector is easy to demonstrate.  
Conditions of continuity and single-valuedness of
wavefunctions are no longer required.
If the circuit surrounds a singularity described
by a $2\times2$ effective Hamiltonian, then
the flux equals half the solid angle $\Omega(C)$ subtended in
an appropriate scaled space by the circuit as seen from the
point of singularity.  The appropriate scaled space is
the one in which the $2 \times 2$ Hamiltonian for states
near the conical intersection has the form ${\cal H}_{\rm eff}=\vec{R}
\cdot\vec{\sigma}$ in terms of scaled coordinates
$\vec{R}=(X,Y,Z)$ and Pauli matrices $\vec{\sigma}=
(\sigma_x,\sigma_y,\sigma_z)$.  This method will be used
twice in this paper.  The eigenvalues
are $\pm R=\rho_z R$, where the quantum number $\rho_z=\pm 1$
is introduced as a branch index.  The geometric phase
is then $\gamma(C)=-\rho_z \Omega(C)/2$.

Mikitik and Sharla\v{i}\cite{Mikitik2007}
provide convincing evidence that the geometric phase $\pm\pi$
is seen experimentally as a shift in the
semiclassical quantization condition \cite{Kosevich}
determining the de Haas-van Alphen oscillations.
An extreme experimental case is the shifted quantum Hall oscillations
originating from orbits near the Dirac points 
in graphene \cite{Graphene1,Graphene2}.

The shifts of quantization condition 
occur for electron orbits (in a $\vec{B}$-field)
which surround a degeneracy line (or point, for graphene.)
Mikitik and Sharla\v{i}  also argue \cite{Mikitik1998} that spin-orbit
effects can mostly be ignored.  This is mostly correct for lighter
elements with spin-orbit strength $\xi/\Delta \ll 1$,
$\Delta$ being any other relevant electron scale such as a band gap.
However, the mathematics and the corrections need elucidation.
Spin-orbit coupling destroys band degeneracy lines, but it
is not evident what happens to the geometric phase of $\pm\pi$.

\begin{figure}
\centerline{\scalebox{0.36}[0.36]{\includegraphics{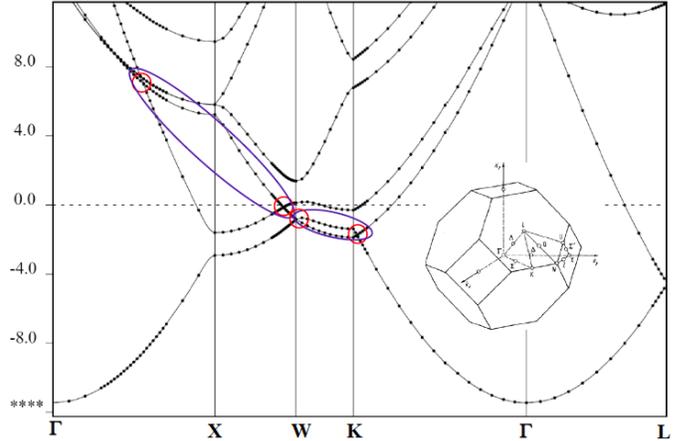}}}
\caption{
The band structure of fcc aluminum, with energy in eV. The red circles 
indicate the important degeneracies, symmetry determined if at a
symmetry point, otherwise accidental. On the Brillouin zone in the
inset, the blue loops show the approximate positions of the nodal
loop, from two viewpoints.
}
\label{fig:Al}
\end{figure}

Mikitik and Sharla\v{i} have shown\cite{Mikitik1999} that, in the neglect of SOC, when
a cyclotron orbit encircles a nodal line, the areal quantization is
shifted, as mentioned earlier. 
This result is a topological one, depending neither on the form of
$\epsilon_k$ nor the size or shape of the orbit. This situation occurs
for certain orbits in fcc aluminum. Nodal loops were mapped in Al
by one of the authors\cite{PBA-unpubl} before the recent wider awareness
of occurrence of nodal loops. The position of one loop near E$_F$ is shown in
Fig.~\ref{fig:Al}. The
locations of the related degeneracies in the electronic spectrum along
symmetry lines are also denoted in
Fig.~\ref{fig:Al}. 
 
\section{Spin-orbit Coupling}
\subsection{Lifting of degeneracy}
To see the effect of spin-orbit interactions, add to ${\cal H}(\vec{k})$
the SOC piece 
\begin{eqnarray}
{\cal H}_{SO}=(\vec{\sigma}/4m^2c^2)\cdot\vec{\nabla}V \times
(\vec{p}+\hbar\vec{k}).
\end{eqnarray}
Choose some point $\vec{k}^{\ast}$ of
accidental degeneracy, and find energies and eigenstates at nearby
$\vec{k}$-points using degenerate $\vec{k}\cdot\vec{p}$ perturbation theory.
For notational simplicity, $\vec{k}^{\ast}$ is 
the temporary origin of $\vec{k}$.  
The degenerate basis functions $|1\rangle$ and $|2\rangle$ are the
periodic parts $u_1$ and $u_2$ at $\vec{k}=\vec{k}^{\ast}$.  
A phase convention is needed; the coefficients $C_G$ of
the expansion $u(\vec{r})=\sum C_G \exp(i\vec{G}\cdot\vec{r})$
are chosen real.  This requires inversion symmetry, which
is hereafter assumed.
Each state has two spin orientations, so the effective 
Hamiltonian matrix is $4\times 4$, with the form
\begin{equation}
{\cal H}_{\rm eff}=\left(\begin{array}{cc}
\hbar\vec{k}\cdot\vec{v}_a\hat{1}  
& \hbar\vec{k}\cdot\vec{v}_b\hat{1} -i\vec{\xi}\cdot\vec{\sigma} \\
\hbar\vec{k}\cdot\vec{v}_b\hat{1} +i\vec{\xi}\cdot\vec{\sigma} 
& -\hbar\vec{k}\cdot\vec{v}_a\hat{1} \end{array}\right)
\label{eq:Heff}
\end{equation}
where $\hat{1}$ and $\vec{\sigma}$ are $2\times 2$ matrices in spin space.
Terms proportional to the $4\times 4$ unit matrix do not mix or split the
states and are omitted.   The vector $\vec{v}_a$ is 
half the relative velocity $(\vec{v}_1 - \vec{v}_2)/2$,
where $\vec{v}_n$ is the band velocity $\vec{\nabla}_k \epsilon_n/\hbar$
at the degeneracy $\vec{k}^{\ast}$.  The vector $\vec{v}_b$ is the off-diagonal
term $\langle 2|\vec{p}/m|1\rangle$, which is pure real 
since $C_G$ is real.  The ``orbital moment'' vector 
\begin{eqnarray}
i\vec{\xi} = \langle2|\vec{\nabla}V
\times\vec{p}|1\rangle/4m^2 c^2
\end{eqnarray}  
is pure imaginary since there is also
time-reversal symmetry, under an assumption of
no magnetic order or external $\vec{B}$-field.  Thus 
three real vectors, $\vec{v}_a$, $\vec{v}_b$, and $\vec{\xi}$, determine
the bands near $\vec{k}^{\ast}$.  
The vector $\vec{\xi}$ is a close analog to angular momentum,
hence the designation as orbital moment.  
Consider a system with two degenerate 
$p$-states $|x\rangle$ and $|y\rangle$.  The angular momentum
operator $\vec{L}$ has an imaginary off-diagonal element.  The mixed states
$|x\rangle\pm i |y\rangle$
are eigenstates of $\vec{L}$ with 
$\langle\vec{L}\rangle=\pm m\hbar\hat{z}$.
The magnitude $m$ deviates from $1$ if the point symmetry is less
than spherical. 

\begin{figure}
\centerline{\scalebox{0.30}[0.30]{\includegraphics{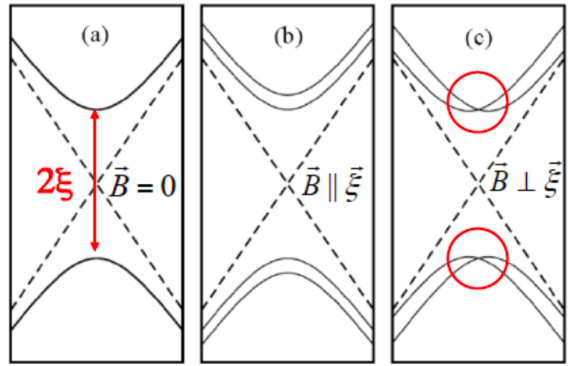}}}
\caption{Energy {\it versus} $|\vec{k}|$ near the degeneracy point,
for (a) no magnetic field, the gap is 2$\xi$, (b) field parallel to $\vec{\xi}$, and
(c) field perpendicular to $\vec{\xi}$, with the degeneracies 
emphasized.  The dashed lines are for $\xi=0$ 
and $b=0$; solid lines in panel (a) are $\pm\lambda$, which becomes
$\pm\xi$ at the degeneracy point $\vec{k}=0$.
}
\label{fig:1}
\end{figure}

First suppose that $\vec{\xi}=0$.  Since $\vec{v}_a$ and $\vec{v}_b$
are not generally collinear and provide the directions along which
matrix elements of ${\cal H}_{\rm eff}$ vary, 
they define a direction of $\vec{k}$, namely
$\vec{v}_a \times \vec{v}_b$, along which ${\cal H}_{\rm eff}=0$.  This is
the {\it direction of the line of degeneracy}.  After allowing $\vec{\xi}\ne 0$,
 eigenvalues of Eq.(\ref{eq:Heff}) are $\pm\lambda$ where
\begin{equation}
\lambda = \sqrt{\kappa_a^2 + \kappa_b^2 + \xi^2}
\label{eq:lambda}
\end{equation}
with $\kappa_a=\hbar\vec{k}\cdot\vec{v}_a$,
$\kappa_b=\hbar\vec{k}\cdot\vec{v}_b$, and $\xi=|\vec{\xi}|$.
Each eigenvalue belongs to a Kramers doublet of two opposite spin states.
The original degeneracy (without spin-orbit interaction) of 2
(neglecting spin) or 4 (including spin) is lifted everywhere unless
$\vec{\xi}=0$.  This should
happen only at isolated points in the Brillouin zone, not
coinciding with degeneracy lines $\vec{k}^{\ast}$.  No accidental
degeneracies remain, but Kramers degeneracy occurs everywhere.  
Bands near $\vec{k}^{\ast}$ are shown in Fig.\ref{fig:1}.

\subsection{Geometric phase}
The geometric phase under consideration involves a circuit $C(\vec{k})$
surrounding the $\vec{k}^{\ast}$ line.  A circular path
in two-dimensional $(\kappa_a,\kappa_b)$-space,
namely $C=(\kappa\cos\phi,\kappa\sin\phi), \ 0\rightarrow\phi
\rightarrow 2\pi$ is the simplest realization.  
To calculate $\gamma(C)$, separate Eq.(\ref{eq:Heff})
into two similar $2\times2$ submatrices by choosing basis states with
spins polarized along $\vec{\xi}$, which will be 
used as the $z$-axis of spin space.  The submatrices are
\begin{equation}
{\cal H}_{{\rm eff}}^{\pm}=\left(\begin{array}{cc}
\kappa_a & \kappa_b \mp i\xi \\
\kappa_b \pm i\xi & -\kappa_a \end{array}\right),
\label{eq:Heffpm}
\end{equation}
where the upper sign goes with spin up, $\sigma_z=1$.  

The circuit can now be considered as
a path $C(\vec{\lambda})$ in a 3-d $\vec{\lambda}$-space,
where $(\lambda_x,\lambda_y,\lambda_z)=(\kappa_b,\sigma_z\xi,\kappa_a)$.
On this circuit, $\lambda$, $\kappa$, and $\xi$ are all constant.
The effective Hamiltonian has the desired scaled form.
The solid angle is $\sigma_z 2\pi(1-\xi/\lambda)$, so the geometric phase is 
\begin{equation}
\gamma(C)=-(\Lambda_z\sigma_z)\pi(1-\xi/\lambda).
\label{eq:zphase}
\end{equation}
where $\Lambda_z=\pm 1$ is the branch index.
This is {\it one of the two main results} of this section.
It shows how spin-orbit splitting destroys the simple phase
of $\pm\pi$ when the circuit has such a small radius that $\xi\sim\lambda$.  
If spin-orbit interaction is weak, it does not need a large orbit to
have $\xi/\lambda \ll 1$ and approach the full simple phase of $\pm\pi$.

\subsection{The  rest of the story}
This is not the full story.  Analogous to the lack of phase $u_k$ not 
being defined by the Schr\"odinger equation, the choice to evolve at fixed
$\sigma_z$ was arbitrary.  The states of Kramers
doublets can be mixed by arbitrary unitary transformations.
Evolution of a doublet around a circuit introduces not a simple
geometric phase, but a unitary matrix.  The $\gamma(C)$ phases
just computed are actually the diagonal elements $\exp(\pm i\gamma(C))$
of a $2\times2$ unitary matrix in the representation with spin
quantized along $\vec{\xi}$.  It will emerge below that this is indeed the 
correct adiabatic evolution of the Kramers doublet when
a small magnetic field is imposed along
the $\vec{\xi}$ direction.

Berry's original argument 
assumed that ${\cal H}$ had a discrete spectrum along {\it C}.
There is a physically natural way to
retain this.  Magnetic fields 
are used to cause cyclic evolution
in $\vec{k}$-space.  Magnetic fields also lift Kramers degeneracy.
The simplest theoretical device is to
add to ${\cal H}_{\rm eff}$ a Zeeman 
term ${\cal H}_Z =-\vec{b}\cdot\vec{\sigma}$
coupling only to spin.

To proceed further,
an explicit representation of eigenstates is needed.
Eigenstates of the effective Hamiltonian (\ref{eq:Heffpm}),
labeled by energy $\pm\lambda$ and $\sigma_z=\uparrow,\downarrow$ are
chosen as
\begin{equation}
|s\rangle=|-\lambda,\uparrow\rangle=\frac{1}{n}\left (\begin{array}{c}
-\kappa_b+i\xi \\ \kappa_a+\lambda \end{array} \right)
\otimes|\uparrow\rangle,
\label{eq:lowerup}
\end{equation}
\begin{equation}
|t\rangle=|-\lambda,\downarrow\rangle=\frac{1}{n}\left(\begin{array}{c}
-\kappa_b-i\xi \\ \kappa_a+\lambda\end{array}\right)
\otimes|\downarrow\rangle,
\label{eq:lowerdown}
\end{equation}
\begin{equation}
|u\rangle=|+\lambda,\uparrow\rangle=\frac{1}{n}\left (\begin{array}{c}
\kappa_a+\lambda \\ \kappa_b+i\xi \end{array} \right)
\otimes|\uparrow\rangle,
\label{eq:upperup}
\end{equation}
\begin{equation}
|v\rangle=|+\lambda,\downarrow\rangle=\frac{1}{n}\left(\begin{array}{c}
\kappa_a+\lambda \\ \kappa_b-i\xi\end{array}\right)
\otimes|\downarrow\rangle.
\label{eq:upperdown}
\end{equation}
These are written as direct product of spatial times spin two-vectors. 
The normalization is $n=\sqrt{2\lambda(\lambda+\kappa_a)}$.
As long as $\xi$ is non-zero, $1/n$ is non-singular and
these are smooth, single-valued functions of
$(\kappa_a,\kappa_b)$, unique except for an arbitrary overall phase,
which cannot alter $\gamma(C)$.
The lower Kramers doublet $|s\rangle,|t\rangle$ has ``orbit moments''
$\langle i|\vec{\nabla}V \times\vec{p}/4m^2 c^2|i\rangle
=\mp(\xi/\lambda)\vec{\xi}$
oriented antiparallel to spin, while the
upper Kramers doublet $|u\rangle,|v\rangle$ has
identical orbit moments except oriented parallel to spin.

Now the Zeeman term is added.
Diamagnetic coupling is neglected.  Without loss of generality,
the part of the field $\vec{B}=\vec{b}/\mu_B$ perpendicular to $\xi$ can
be used to define the $x$ direction of spin.
The total Hamiltonian in the basis 
$|s\rangle,|t\rangle,|u\rangle,|v\rangle$ is
\begin{equation}
{\cal H}_{\rm tot}=-\left(\begin{array}{cccc}
\lambda+b_z&\frac{\kappa}{\lambda}b_x e^{i\omega}&0&i\frac{\xi}{\lambda}b_x\\
\frac{\kappa}{\lambda}b_x e^{-i\omega}&\lambda-b_z&-i\frac{\xi}{\lambda}b_x&0\\
0&i\frac{\xi}{\lambda}b_x&-\lambda+b_z&\frac{\kappa}{\lambda}b_x e^{-i\omega}\\
-i\frac{\xi}{\lambda}b_x&0&\frac{\kappa}{\lambda}b_x e^{i\omega}&-\lambda-b_z
\end{array}\right)
\label{eq:Htot}
\end{equation}
The factor $(\kappa/\lambda) \exp(i\omega)=\langle s|\sigma_+|t\rangle$
introduces the new angle $\omega$ 
\begin{equation}
e^{i\omega}=\frac{\lambda}{\kappa}-\frac{\xi(\xi-i\kappa_b)}{\kappa(\lambda+\kappa_a)}.
\label{eq:omega}
\end{equation}
As the circuit {\it C} is followed ($\phi$ going from 0 to $2\pi$,
with $\xi,\kappa,\lambda$ constant),
$\omega$ also evolves from 0 to $2\pi$.

If the field $\vec{b}$ is along $z$, the upper and lower Kramers
doublets are not coupled.  The degeneracy is lifted everywhere,
and adiabatic evolution proceeds smoothly on the resulting non-degenerate
states, yielding the phases $\gamma(C)$ of Eq.(\ref{eq:zphase}).  
The previous discussion was correct.
The result Eq.~\ref{eq:zphase} can also be obtained directly from
Eq.(\ref{eq:geomphase}) using 
Eqs.(\ref{eq:lowerup},\ref{eq:lowerdown},\ref{eq:upperup},\ref{eq:upperdown}).
For fields perpendicular to $z$, there is both intra- and
inter-doublet spin mixing, according to Eq.(\ref{eq:Htot}).
To first order, since $\vec{b}\ll\lambda$,
inter-doublet mixing terms $\pm i\xi b_x /\lambda$ can
be neglected, giving $2\times2$ effective Hamiltonian
matrices, of the form
\begin{equation}
{\cal H}_{\rm eff}(\vec{b}) =\lambda_z \lambda\hat{1} 
- \left( \begin{array}{cc}
b_z & \frac{\kappa}{\lambda}b_x e^{i\lambda_z \omega} \\
\frac{\kappa}{\lambda}b_x e^{-i\lambda_z \omega} & -b_z
\end{array}\right)
\label{eq:Heffb}
\end{equation}
The eigenvalues are
\begin{equation}
\pm \lambda \pm \mu \ \ \ {\rm where} \ \mu^2=b_z^2 + \frac{\kappa^2}
{\lambda^2}b_x^2
\label{eq:beigenvalues}
\end{equation}
These eigenvalues have an interesting feature: 
at the degeneracy point $\kappa=0$, in the center of circuit {\it C},
$\mu=0$ and Kramers degeneracy is {\bf not} lifted,
provided $\vec{b}$ is
perpendicular to $\vec{\xi}$.  The states at $\vec{k}^{\ast}$
have anisotropic $g$ factors which vanish in two directions.
The vanishing Zeeman splitting means 
that a conical intersection, hidden unless $\vec{b}\ne 0$, 
exists exactly where the original band
intersection (for $\xi=0$) was located.  This also yields a simple
geometrical phase of $\pm\pi$.  Bands for $\vec{b}\parallel\vec{\xi}$
and $\vec{b}\perp\vec{\xi}$ are shown in Fig.\ref{fig:1} panels (b)
and (c).

A full calculation of $\gamma(C)$ for the 4 new eigenstates
of Eq.(\ref{eq:Htot}) is difficult.  
The Berry method of solid angle works when the basis
functions $|1\rangle$, $|2\rangle$ of the $2\times2$
effective Hamiltonian are fixed at $\vec{k}^{\ast}$,
whereas the basis functions $|s\rangle,|t\rangle$ or
$|u\rangle,|v\rangle$ used in Eq.(\ref{eq:Heffb}) depend
on $\vec{k}$.  However, the most important limit
remaining to be resolved is when the circuit radius $\kappa$
is small relative to spin-orbit splitting $\xi$.  In this limit,
the basis functions lose their $\vec{k}$-dependence. 
The relevant scaled parameters are $\vec{\mu}=((\kappa/\lambda)
b_x\cos\omega,-\lambda_z(\kappa/\lambda)b_x\sin\omega,b_z)$.
The circuit parameterized by $\phi$ is equally well
parameterized by $\omega$ which evolves from 0 to $2\pi$.
The solid angle in $\vec{\mu}$-space is $2\pi\Lambda_z(1-b_z/\mu)$,
so the geometric phase is
\begin{equation}
\gamma(C)=-\pi\beta_z\Lambda_z\left(1-\frac{b_z}{\sqrt{b_z^2+
(\kappa/\lambda)^2 b_x^2}}\right),
\label{eq:finalphase}
\end{equation}
where $(\Lambda_z,\beta_z)$ are the two branch indices in the
eigenvalue $\pm\lambda\pm\mu=\Lambda_z\lambda+\beta_z\mu$.  
This is the {\it other main result} of this section.
If $b_z=0$, the full
geometric phase $\gamma(C)=\pm\pi$ is restored no matter
how small the circuit radius.  Even though the degeneracy
was lifted by spin-orbit interactions, the hidden conical
intersection exposed by a Zeeman field controls the result.
The nodal loop can be located and followed, though it never
existed.

\section{Connections to the Lowest Symmetry Nodal Line Semimetal}
Reports of identification of nodal loops in electronic structures took
off in 2014. There had been an early report in 2009, 
before widespread recognition
of nodal loops resulted from the 2011 paper of  Burkov, Hook, and
Balents.\cite{Burkov2011} These authors popularized nodal loop semimetals 
in the context of
topological semimetals (primarily Weyl semimetals),
pointing out several general features. 
The earlier report of Pardo and Pickett\cite{Pardo2010}
involved a nodal loop comprised of a pair of 
coinciding Fermi rings, making it a circular nodal ring coinciding with
the Fermi energy $E_F$, a simple but remarkable coincidence. The system was a
compensated semimetal of ferromagnetic nanolayers of SrVO$_3$ 
quantum confined within insulating 
SrTiO$_3$. Mirror symmetry was a central feature: two bands
having opposite reflection symmetries crossed in the mirror plane, making it a
nodal loop enabled by symmetry (thus not purely accidental). 

What is unlikely but not statistically improbable is: (1) having the nodal loop cut by
$E_F$  while (2) the remainder of the Brillouin zone is gapped. Such loops will
have real impact, and possibly unusual boundary properties, when they are the sole bands around
$E_F$. 
This coincidence with E$_F$ occurred
for the ferromagnetic SrVO$_3$ nanolayer mentioned above.
Crystal symmetry has played an important role in nearly all nodal loop
families discovered so far. The enabling symmetries include screw 
axes,\cite{Fang2015}   
mirror symmetries,\cite{Pardo2010,Burkov2011,Phillips2014,Fang2015,Heikkila2015,Mullen2015,Fang2012}  
as well as the much studied TaAs class that has no center of 
inversion.\cite{Huang2015,Xu2015,Lv2015,Shekhar2015,Weng2015,Ahn2015,Sun2015,Yang2015}

Herring\cite{Herring1937, HerringThesis}
however pointed out that inversion symmetry ${\cal P}$ alone is 
sufficient to allow nodal loops of degeneracies 
(fourfold: two bands times two spins), a result extended 
recently.\cite{Burkov2011,Fang2015} 
This is easy to understand:  ${\cal P}$ symmetry leads to a real 
Bloch Hamiltonian $H(\vec k)$ if the 
center of inversion is taken as the origin. The minimal 
(for each spin) 2$\times$2 Hamiltonian 
then has the form $H(\vec k)=f_k\tau_x + g_k\tau_z$ 
(neglecting  spin degeneracy for the moment) 
with real functions $f_k, g_k$; $\vec \tau$ represents the Pauli matrices in orbital space. 
Degeneracy of the eigenvalues $\varepsilon_k = \pm(f_k^2 + g_k^2)^{1/2}$ 
requires $f_k=0=g_k$, two conditions 
on the 3D vector $\vec k$ giving the necessary flexibility to arrange degeneracy.
Allen has given a constructive
prescription\cite{Allen2007} for mapping the nodal loop once a degeneracy is located.

The Zintl semimetal CaAs$_3$, which has $P{\bar 1}$ symmetry (inversion only) 
 has been shown\cite{CaAs3} to have, before SOC is considered, a nodal loop that is cut by the
Fermi energy four times. It and its three isovalent 
tri-arsenide sisters (Ca$\rightarrow$Sr, Ba, Eu) were synthesized more 
than thirty years ago, with their structure, transport, and optical properties studied by 
von Schnering, Bauhofer, and collaborators.\cite{bauhofer1981,Oles1981} 

CaAs$_3$ is unique in a few ways. It sports a single nodal loop. Already Herring had
noted that three classes were possible: single nodal loops, loops that occur in pairs,
and loops that are extended into neighboring zones, being ``closed'' by the periodicity
of $k$ space. Other than these tri-arsenides, nodal loop semiconductors all have
pairs of nodal loops imposed by their crystal symmetry.
CaAs$_3$ also is the sole triclinic ($P{\bar 1}$) member of this family
of tri-arsenides.\cite{bauhofer1981} 
CaAs$_3$ also has the accidental feature that the SOC splitting of the nodal
loop $\Delta E_{soc}$ (arising from the As SOC) is very similar to its small 
dispersion of 30-40 meV around the loop.
This similarity of energy scales leaves the resulting band structure on the
borderline between remaining a nodal loop semiconductor or moving into the realm
of extremely small gap nodal insulator; this distinction is too small for present
DFT calculations to give conclusive statements about. Recall also that such small
gap systems are unstable to excitonic condensation. 

Thus CaAs$_3$ presents a unique nodal semimetal amongst those discovered and
studied so far. Unfortunately, CaAs$_{3}$ samples
are heavily twinned due to a structural transition between the growth temperature and
the temperatures of interest (room temperature and below). The twin boundaries
likely produce carriers that will complicate interpretation of transport and
spectroscopic data.  
The results of Sec. III point out the conceivability of identifying the nodal loop
even though it has been destroyed by spin-orbit coupling. The experimental challenge is
constructing and experimental realization of the theoretical ``SQUID loop'' -- the
circuit {\cal C} -- that enables detection of the loop of degeneracies. This possibility
provides impetus for discovering, or designing, other nodal loop semimetals with
minimal symmetry.

\section{Acknowledgments}
P.B.A. thanks A. G. Abanov and M. S. Hybertsen for assistance,
and also thanks the Stony Brook students of 2007 Phy556 who were subjected to preliminary
versions of the formalism presented in this work. W.E.P. acknowledges
collaboration with Y. Quan on CaAs$_3$.\cite{CaAs3}
P.B.A was supported earlier by NSF grant no.
NIRT-0304122 and currently by DOE grant DE-FG02-08ER46550.
W.E.P. was supported by DOE grant DE-FG02-04ER46111. 

\end{document}